# Dependence of single-particle energies on coupling constants of the nuclear energy density functional

M. Kortelainen,[1] J. Dobaczewski,[1,2] K. Mizuyama,[1] and J. Toivanen[1]

[1]*Department of Physics, P.O. Box 35 (YFL), FI-40014 University of Jyväskylä, Finland*
[2]*Institute of Theoretical Physics, University of Warsaw, ul. Hoża 69, 00-681 Warsaw, Poland.*
(Dated: November 2, 2018)

We show that single-particle energies in doubly magic nuclei depend almost linearly on the coupling constants of the nuclear energy density functional. Therefore, they can be very well characterized by the linear regression coefficients, which we calculate for the coupling constants of the standard Skyrme functional. We then use these regression coefficients to refit the coupling constants to experimental values of single-particle energies. We show that the obtained rms deviations from experimental data are still quite large, of the order of 1.1 MeV. This suggests that the current standard form of the Skyrme functional cannot ensure spectroscopic-quality description of single-particle energies, and that extensions of this form are very much required.

## I. INTRODUCTION

The energy-density-functional (EDF) methods are recently intensely studied in nuclear physics to gain more precise and predictive description of stable and exotic nuclei within a unified framework. Such methods can be considered as realizations of the nuclear effective theory at low energies [1] and are motivated by successful applications of the density functional theory in electronic systems, see, e.g., Ref. [2]. In nuclear physics, they have been studied since early 1970s under the names of Hartree-Fock or Hartree-Fock-Bogolyubov methods [3], but in fact they rather correspond to the Kohn-Sham [4] type of approaches, aiming at describing correlated fermion systems by their one-body (local or non-local) densities only.

In the present study we focus on the Skyrme-type EDFs, by which one assumes that the ground-state energy of a given nucleus is given by an integral of a local energy density. As discussed in Ref. [5], our main goal is to look for the spectroscopic-quality EDF, which would correctly describe positions of single-particle (s.p.) energies across the nuclear mass table. To this end, here we analyze dependence of the s.p. energies on the EDF coupling constants and attempt answering the question on whether the current parametrizations of the EDF are rich enough to describe experimental data with reasonable precision. In the present work we look only at bare s.p. energies in doubly magic nuclei, because the polarization effects, which may affect detailed values of s.p. energies, are estimated to be significantly smaller than discrepancies with experimental data [5].

The paper is organized in the following way. In Sec. II we briefly introduce necessary definitions and present our method of analysis. Three subsections of Sec. III present our results on s.p. energies in function of coupling constants, regression coefficients, and fits to experimental data. Conclusions of our work are presented in Sec. IV.

## II. METHOD

In the present study we consider EDF in the form given in Refs. [6, 7],

$$\mathcal{E} = \int d^3 r \mathcal{H}(\boldsymbol{r}), \quad (1)$$

where the energy density $\mathcal{H}(\boldsymbol{r})$ can be represented as a sum of the kinetic energy and of the potential-energy isoscalar ($t=0$) and isovector ($t=1$) terms,

$$\mathcal{H}(\boldsymbol{r}) = \frac{\hbar^2}{2m}\tau_0 + \mathcal{H}_0(\boldsymbol{r}) + \mathcal{H}_1(\boldsymbol{r}), \quad (2)$$

which for the time-reversal and spherical symmetries imposed read:

$$\mathcal{H}_t(\boldsymbol{r}) = C_t^\rho \rho_t^2 + C_t^{\Delta\rho}\rho_t \Delta \rho_t + C_t^\tau \rho_t \tau_t + \tfrac{1}{2} C_t^J \boldsymbol{J}_t^2 + C_t^{\nabla J} \rho_t \boldsymbol{\nabla}\cdot\boldsymbol{J}_t. \quad (3)$$

Standard definitions of the local densities $\rho_t$, $\tau_t$, and $\boldsymbol{J}_t$ were given in Refs. [7, 8] and are not repeated here. Following the parametrization used for the Skyrme forces, we assume the dependence of the coupling parameters $C_t^\rho$ on the isoscalar density $\rho_0$ as:

$$C_t^\rho = C_{t0}^\rho + C_{tD}^\rho \rho_0^\alpha. \quad (4)$$

Similarly as in Ref. [9], we note here that EDF of Eq. (1) depends *linearly* on twelve coupling constants,

$$C_{t0}^\rho, \quad C_{tD}^\rho, \quad C_t^\tau, \quad C_t^{\Delta\rho}, \quad C_t^{\nabla J}, \quad \text{and} \quad C_t^J, \quad (5)$$

for $t=0$ and 1. Therefore, due to the Hellmann-Feynman theorem [10], derivatives of the total energy $\mathcal{E}$ with respect to the coupling constants are given by space integrals of densities appearing in Eq. (3) [9].

Variation of EDF in Eq. (1) with respect to the Kohn-Sham orbitals $\phi_i(\boldsymbol{r})$, which define the local densities, gives the standard eigenequation,

$$\hat{h}\phi_i = \epsilon_i \phi_i, \quad (6)$$

where the Kohn-Sham one-body Hamiltonian $\hat{h}$ is obtained as a functional derivative of EDF with respect to densities, see, e.g., Refs.[7, 8, 11]. It is obvious that $\hat{h}$ also depends linearly on the coupling constants, and therefore, again due to the Hellmann-Feynman theorem, derivatives of the s.p. energies $\epsilon_i$ with respect to the coupling constants are also given by space integrals of densities. As a consequence, we can expect that these derivatives are generic quantities weakly dependent on a particular parametrization of the functional, provided the functional has been adjusted to data and the underlying densities are correct.

The aim of the present work is not only to determine these derivatives but also to estimate to which extent they are generic. Of course, for a given set of coupling constants, they can be calculated from the Hellmann-Feynman theorem. Equivalently, their determination may rely on numerically calculating functions $\epsilon_i(C_m)$, where index $m = 1, \ldots 12$ is assumed to enumerate the twelve coupling constants (5). One simple option would be to calculate them from the finite-difference formula, for example as

$$\frac{\partial \epsilon_i}{\partial C_m}(C_m^0) \simeq \frac{\epsilon_i(C_m^+) - \epsilon_i(C_m^-)}{C_m^+ - C_m^-}, \qquad (7)$$

with values of $C_m^+$ and $C_m^-$ suitably close to $C_m^0$. Had the functions $\epsilon_i(C_m)$ been exactly linear, the Hellmann-Feynman and finite-difference methods would have given exactly the same answers. If we aim at determining the degree to which they are not linear we have to proceed in another way.

To this end, the method of choice is the linear regression analysis [12], which makes a hypothesis of linearity and tests it by determining the regression coefficients and their standard deviations. In our case, we write the expression

$$\epsilon_i(C_m^0 + d_m^k) = I_{im} d_m^k + I_i^0 + r_i^k, \qquad (8)$$

where $\epsilon_i(C_m^0 + d_m^k)$ are the s.p. energies calculated self-consistently for coupling constants $C_m^0 + d_m^k$ that differ from $C_m^0$ by suitably small numbers $d_m^k$. Index $k$ enumerates different choices of these small numbers (which sample the vicinity of $C_m^0$) and $r_i^k$ are the residuals between the self-consistent results and linear approximation given by the regression coefficients $I_{im}$ and $\epsilon_i^0$. The regression method minimizes the residuals by adjusting the regression coefficients and determines their standard deviations. It is obvious that regression coefficients $I_{im}$ constitute estimates of derivatives (7), while $I_i^0$ are very close to $\epsilon_i(C_m^0)$.

The results below were obtained by using $d_m^k = \pm d C_m^0$, where $d = 0.001$ to $0.005$, depending on the Skyrme force and nucleus, i.e., each of the twelve coupling constants was raised or lowered by the same percent fraction. (In some cases, for vanishing coupling constants $C_m^0$, appropriate absolute shifts were used.) As a result, in each case the regression analysis was done by using $k = 1, \ldots, 2^{12} = 4096$ samples.

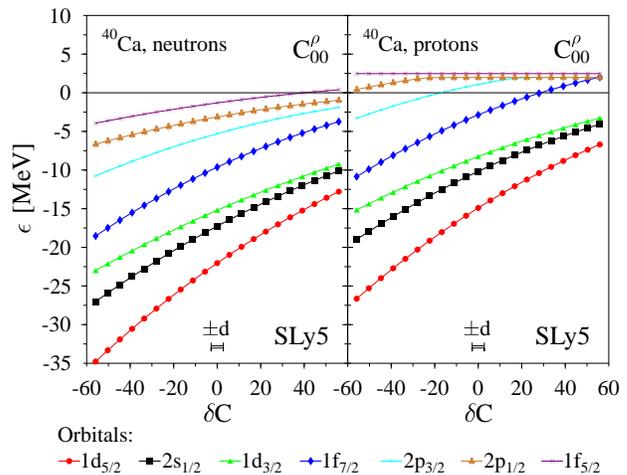

FIG. 1: (Color online) Neutron (left) and proton (right) s.p. levels in $^{40}$Ca in function of $\delta C = C_{00}^\rho - C_{00}^\rho(\text{SLy5})$, where $C_{00}^\rho(\text{SLy5}) = -931.294$ MeV fm$^3$ (as well as other, unchanged, coupling constants) correspond to the Skyrme force SLy5. Symbols listed in the legend from left to right correspond to increasing s.p. energies. The range of the $d$ parameter used in the regression analysis is also indicated.

## III. RESULTS

In the present work, the effect of the EDF coupling constants on s.p. energies was studied in doubly magic nuclei. Calculations were performed by using the code HFBRAD [13], which solves the self-consistent equations with spherical symmetry assumed. Nuclei considered in our calculations were $^4$He, $^{16}$O, $^{40,48}$Ca, $^{48,56}$Ni, $^{100,132}$Sn, and $^{208}$Pb. The coupling constants $C_m^0$ (5) corresponding to the Skyrme functionals SLy5 [14], SkP [15], and SkO' [16] were used. In order to consistently calculate the energies of the occupied and unoccupied states, the program was run in the HFB mode. Even if the pairing correlations vanish in doubly-magic nuclei, a residual non-zero pairing was kept and equivalent s.p. energies [15] were determined.

### A. Single-particle energies in function of coupling constants

In Figs. 1–5 we show the s.p. energies in $^{40}$Ca as functions of five isoscalar coupling constants (5). Results were obtained by varying one coupling constant of the SLy5 functional and by keeping all the remaining ones at their SLy5 values. Ranges of variation of the coupling constants were chosen in a maximum possible way, i.e., up to the values where changes of level ordering or levels becoming unbound precluded obtaining meaningful solutions. Of course, physical values of these coupling constants are fairly well fixed by adjustments to empirical data, so, in practice, physically acceptable variations must not be so large. Therefore, values of parameters $d$ used in the regression analysis below, also shown in Figs. 1–5, were quite small. Positive s.p. energies are shown only to indicate in which regions of parameters the levels become unbound – precise values of these s.p. energies are only loosely related to positions of resonances.

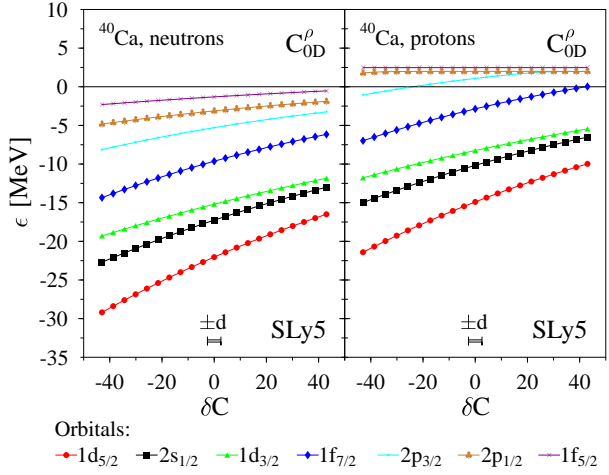

FIG. 2: (Color online) Same as in Fig. 1 but for $\delta C = C_{0D}^\rho - C_{0D}^\rho(\text{SLy5})$, where $C_{0D}^\rho(\text{SLy5}) = 859.813\,\text{MeV fm}^{3(1+\alpha)}$.

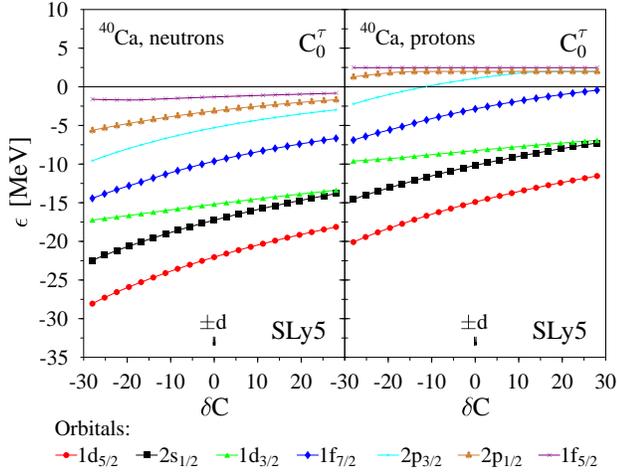

FIG. 3: (Color online) Same as in Fig. 1 but for $\delta C = C_0^\tau - C_0^\tau(\text{SLy5})$, where $C_0^\tau(\text{SLy5}) = 56\,\text{MeV fm}^5$.

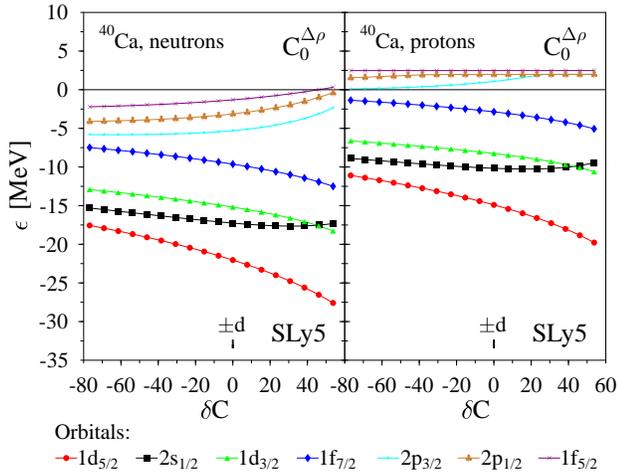

FIG. 4: (Color online) Same as in Fig. 1 but for $\delta C = C_0^{\Delta\rho} - C_0^{\Delta\rho}(\text{SLy5})$, where $C_0^{\Delta\rho}(\text{SLy5}) = -76.793\,\text{MeV fm}^5$.

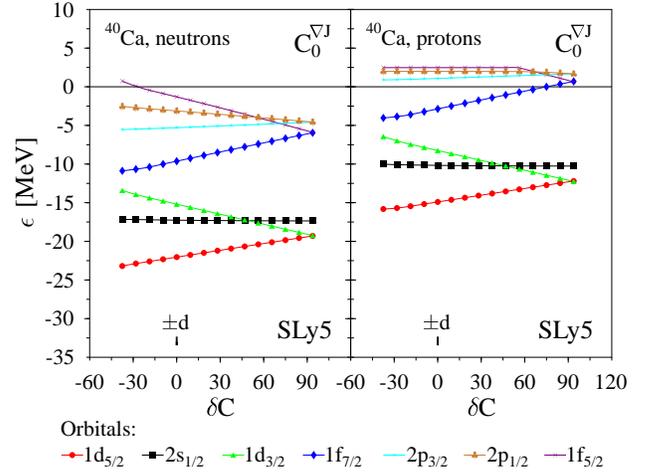

FIG. 5: (Color online) Same as in Fig. 1 but for $\delta C = C_0^{\nabla J} - C_0^{\nabla J}(\text{SLy5})$, where $C_0^{\nabla J}(\text{SLy5}) = -93.75\,\text{MeV fm}^5$.

In Figs. 1–5 one can clearly see that there is only a rather small overall non-linearity of the s.p. energies as functions of the coupling constants. With decreasing values of coupling constants $C_{00}^\rho$ and $C_{0D}^\rho$, the s.p. potentials become deeper and thus the values of s.p. energies uniformly decrease (Figs. 1 and 2). There are only very small differences in the dependencies induced by varying coupling constants $C_{00}^\rho$ and $C_{0D}^\rho$. Somewhat larger changes in relative positions of s.p. levels are induced by varying coupling constant $C_0^\tau$ (Fig. 3). Decreasing values of this coupling constant induce deeper s.p. potentials and larger values of the effective mass. Coupling constant $C_0^{\Delta\rho}$ clearly influences the surface properties of the s.p. potentials (Fig. 4) by changing relative positions of the low-$\ell$ and high-$\ell$ levels. Finally, the SO coupling constant $C_0^{\nabla J}$ very linearly changes the SO splitting of levels shown in Fig. 5. Dependence of levels on the tensor coupling constant $C_0^J$ is very weak in a spin-saturated nucleus $^{40}$Ca, and, therefore, it is not shown.

### B. Regression coefficients

It is clear that results shown in Figs. 1–5 can only tentatively indicate the type of dependence of the s.p. energies on coupling constants. In order to have a better handle on this dependence in a multi-dimensional space of parameters, it is more efficient to directly consider the regression coefficients discussed in Sec. II. For a better separation of central and SO effects, the regression analysis was performed independently for centroids and SO

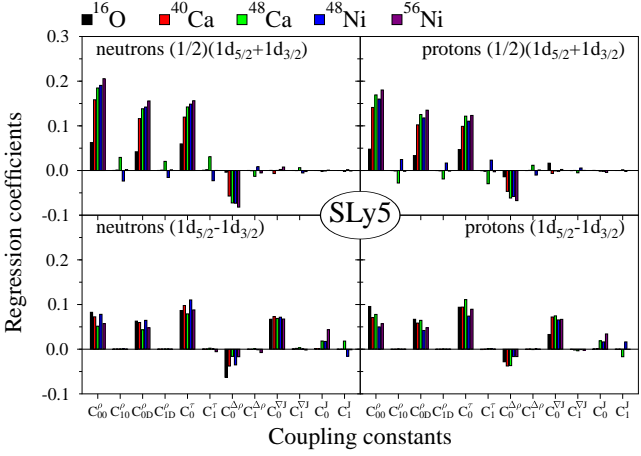

FIG. 6: (Color online) Bar charts of regression coefficients $I_{im}$, Eq. (8), for neutron (left) and proton (right) centroids (top) and SO splittings (bottom), determined for the 2d orbital and SLy5 Skyrme functional. The order of bars is the same as the order of nuclei in the legend on top of the Figure. In order to better compare results for the centroids and SO splittings, regression coefficients are shown for inverted (negative) SO splittings. Units of regression coefficients $I_{im}$ are equal to a MeV divided by the units of coupling constants, which are given in captions of Figs. 1–5.

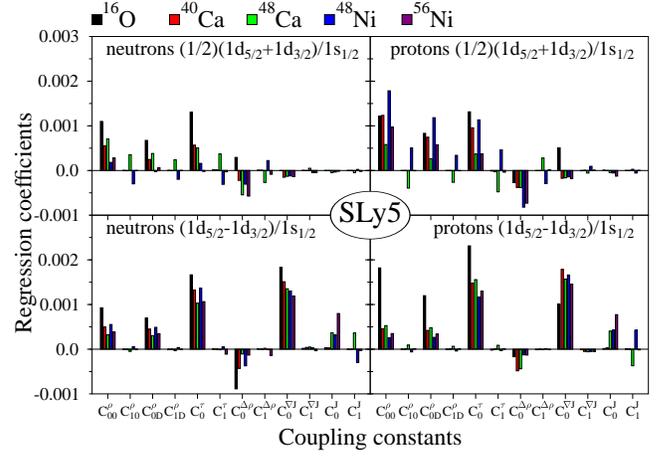

FIG. 7: (Color online) Same as in Fig. 6 but for the regression coefficients $I'_{im}$, Eq. (11).

splittings of the s.p. levels,

$$\epsilon_{n\ell}^{\text{cent}} = \tfrac{1}{2}\left(\epsilon_{n\ell j_<} + \epsilon_{n\ell j_>}\right), \quad (9)$$
$$\epsilon_{n\ell}^{\text{SO}} = \epsilon_{n\ell j_<} - \epsilon_{n\ell j_>}, \quad (10)$$

respectively. Below we present selected results illustrating the main conclusions of the present study.

Regression coefficients obtained for the centroids and SO splittings of the 1d levels in five light nuclei are shown in Fig. 6. As expected, in the $N=Z$ nuclei $^{16}$O, $^{40}$Ca, and $^{56}$Ni, the s.p. energies depend only on the isoscalar coupling constants. Only the centroids in $N\neq Z$ nuclei $^{48}$Ca and $^{48}$Ni weakly depend on the isovector coupling constants $C_{10}^\rho$, $C_{1D}^\rho$, and $C_1^\tau$, and the SO splittings in these nuclei weakly depend on the isovector tensor coupling constant $C_1^J$. In all cases, dependence on the isovector SO coupling constant $C_1^{\nabla J}$ is extremely weak. This illustrates the fact that the isospin excess in $^{48}$Ca and $^{48}$Ni is still not large enough for a pronounced dependence on the isovector coupling constants, and thus these coupling constants can be reasonably fixed only by going to even more exotic nuclei.

Centroids of the 1d levels depend predominantly on four isoscalar coupling constants, $C_{00}^\rho$, $C_{0D}^\rho$, $C_0^\tau$, and $C_0^{\Delta\rho}$. These coupling constants act on positions of centroids in a fairly similar way, i.e., by looking at positions of 1d orbitals in different nuclei one is not able to distinguish between central, effective mass, and surface effects, or differentiate between the density-dependent and density-independent coupling constants.

In all studied nuclei we obtain a standard uniform dependence of the SO splittings on the isoscalar SO coupling constant $C_0^{\nabla J}$. A somewhat weaker, but non-negligible, dependence on the isoscalar tensor coupling constant $C_0^J$ is also clearly seen in spin-non-saturated nuclei $^{48}$Ca, $^{48}$Ni, and $^{56}$Ni. An unexpected result of our study is the fact that the SO splittings also very strongly depend on the central, $C_{00}^\rho$ and $C_{0D}^\rho$, effective-mass, $C_0^\tau$, and surface, $C_0^{\Delta\rho}$ coupling constants. These dependencies were not explicitly recognized in the literature, because most often the SO splittings were adjusted at the end of the fitting protocol, i.e., for fixed values of other coupling constants. It is clear, however, that a global adjustment will create specific inter-relations between the SO and tensor coupling constants on one side and all the other ones on the other side. Relations of these kind between the SO and effective-mass effects were recently analyzed in Ref. [17].

In order to elucidate origins of the obtained dependencies of the SO splittings on the coupling constants, in Fig. 7 we show results of the regression analysis performed for the s.p. energies relative to those of the 1s orbital, i.e., in analogy with Eq. (8),

$$\frac{\epsilon_i(C_m^0+d_m^k)}{\epsilon_{1s}(C_m^0+d_m^k)} = I'_{im}d_m^k + I'^0_i + r_i^k. \quad (11)$$

In this way, we remove the overall effects of scaling of the s.p. spectra related to changing the depth of the central potential and/or effective mass.

One can see that, indeed, in this changed representation, dependence of relative centroids on coupling constants becomes significantly weaker. Also dependence of the relative SO splittings on the central, $C_{00}^\rho$ and $C_{0D}^\rho$, and surface, $C_0^{\Delta\rho}$, coupling constants becomes less pronounced. Nevertheless, the relative SO splittings still quite strongly depend on the effective-mass coupling constant, $C_0^\tau$, i.e., this dependence cannot be attributed to a simple scaling of the s.p. spectrum.

This fact is already clearly visible in Fig. 3, where the s.p. energies of favored SO partners, $1d_{5/2}$ and $1f_{7/2}$,

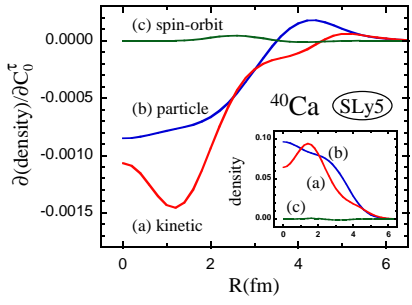

FIG. 8: (Color online) Derivatives of the particle (in fm$^{-3}$), kinetic (in fm$^{-5}$), and spin-orbit (in fm$^{-4}$) neutron densities with respect to the coupling constant $C_0^\tau$ (in MeV fm$^5$). The inset shows the densities themselves. Calculations were performed in $^{40}$Ca for the SLy5 functional.

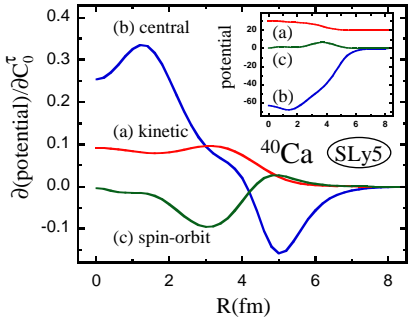

FIG. 9: (Color online) Same as in Fig. 8 but for the central (in MeV), spin-orbit (in MeV fm), and kinetic (in MeV fm$^2$) potentials.

clearly follow that of the $2s_{1/2}$ level, while those of the unfavored SO partners, $1d_{3/2}$ and $1f_{5/2}$, behave quite differently.

The influence of the isoscalar effective-mass coupling constant on the SO splittings can be understood by inspecting the self-consistent densities and potentials. In Fig. 8 we show derivatives of the kinetic (a), particle (b), and SO (c) neutron densities in $^{40}$Ca with respect to the coupling constant $C_0^\tau$. For completeness, the inset shows the total densities. The derivatives were determined from differences of densities calculated self-consistently at two values close to the SLy5 value of $C_0^\tau$. One can see that an increase in $C_0^\tau$ not only lowers the kinetic density $\tau_n$, but also significantly lowers the particle density $\rho_n$ in the center of the nucleus and moves the particles to the surface. As a result, it induces a smaller gradient of the particle density at the surface, and hence a weaker SO potential, as shown in Fig. 9, and as a consequence – a smaller SO splitting.

One can say that the inter-relations between the effective-mass and SO effects stem from the fact that EDF of Eq. (3) contains the effective-mass term, $C_0^\tau \rho_0 \tau_0$, that depends on the product of the particle and kinetic densities. Therefore, changes in the effective-mass coupling constant induce changes in both these densities, and, in particular, they induce changes in the gradient of the particle density that affect the SO splittings. The role of the effective-mass term is clearly not restricted to changing the overall density of levels, as is the case in a trivial case of the infinite square-well potential, cf. also discussion in Ref. [17].

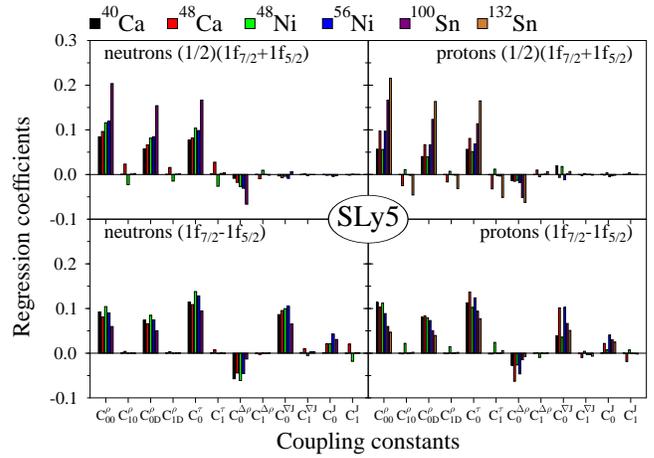

FIG. 10: (Color online) Same as in Fig. 6 but for the 1f orbital.

FIG. 11: (Color online) Same as in Fig. 6 but for the 1g orbital.

Conclusions drawn from the results obtained for the 1d levels are further corroborated by those for the 1f, 1g, and 1h levels, which are shown in Figs. 10, 11, and 12, respectively. One can see that a tangible influence of the isovector coupling constants on centroids is obtained only in $^{132}$Sn and $^{208}$Pb, i.e., in systems with a significant neutron excess. But even there the influence of the isovector coupling constants on the SO splittings is quite weak. Again we see that adjustment of the isovector coupling constants to empirical data may require studying systems much further away from stability.

A universality of the regression coefficients $I_{im}$, Eq. (8), is illustrated in Figs. 13 and 14, in which we present results obtained for the SkP and SkO' Skyrme functionals, respectively. Comparison with the SLy5 results previously presented in Fig. 6, shows that for all the three forces, patterns of regressions coefficients are very similar indeed. Small differences in the overall magnitudes are probably related to differences in the power $\alpha$ of the density dependence, Eq. (4), which is equal to 1/6 (1/4) for SLy5 and SkP (SkO') and in the values of the effective mass, $m^*/m = 0.7$, 0.9, and 1 for SLy5, SkO', and SkP, respectively.

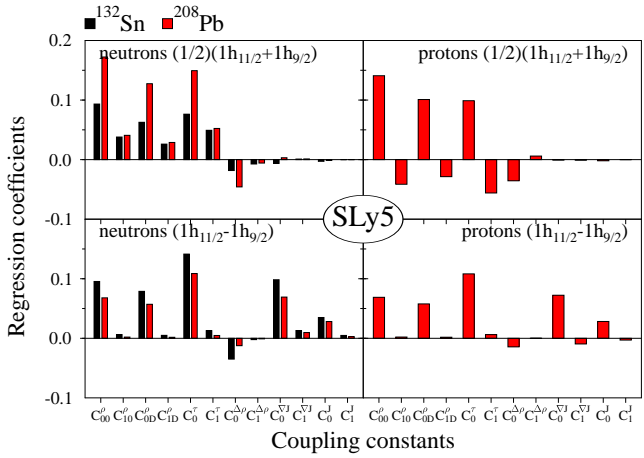

FIG. 12: (Color online) Same as in Fig. 6 but for the 1h orbital.

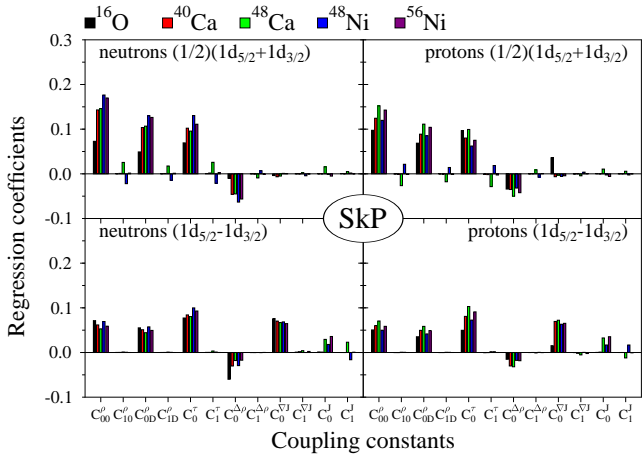

FIG. 13: (Color online) Same as in Fig. 6 but for the SkP Skyrme functional.

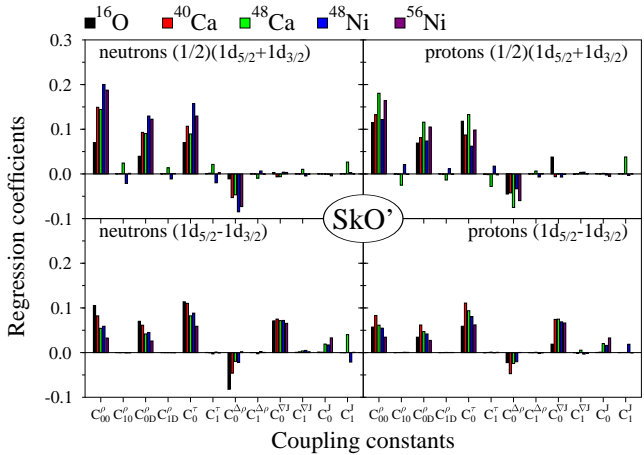

FIG. 14: (Color online) Same as in Fig. 6 but for the SkO' Skyrme functional.

### C. Fits to experimental data

The regression coefficients $I_{im}$ obtained in the present work can be used to evaluate the possibility of readjusting the EDF coupling constants so as to better describe the experimental data. Indeed, if for a given set of coupling constants $C_m$ the self-consistent s.p. energies are equal to $\epsilon_i$, then corrections to coupling constants $\Delta C_m = C_m^{\text{fit}} - C_m$ are given by:

$$\Delta \epsilon_i = \epsilon_i - \epsilon_i^{\text{exp}} = -\sum_m I_{im} \Delta C_m. \qquad (12)$$

The matrix $I_{im}$ being rectangular, it cannot be simply inverted, but, instead, one can use its singular value decomposition [18] to determine the result, i.e.,

$$I_{im} = \sum_n U_{in} w_n V_{nm}^T, \qquad (13)$$

where columns of matrices $U$ and $V$ are orthogonal and singular values $w_n$ are positive. In our case, the number of experimental data $\epsilon_i^{\text{exp}}$ is much larger than the number of coupling constants $C_m$, which is equal to 12. Therefore, the possibility of describing experimental data clearly depends on whether the column $\Delta \epsilon_i$ can be expressed as a linear combination of 12 orthogonal columns of matrix $U$. In any case, the result of the least-square fit to experimental data is given by

$$\Delta C_m = -\sum_{n=1}^{n_{\max}} V_{mn} w_n^{-1} \sum_i U_{ni}^T \Delta \epsilon_i, \qquad (14)$$

for $n_{\max}=12$.

Singular values $w_n$ obtained for regression coefficients $I_{im}$ corresponding to three Skyrme functionals are shown in the lower panel of Fig. 15. Independently of which functional is concerned, the singular values decrease exponentially from about 1.3 to 0.001. Again, this illustrates a great universality of the regression coefficients.

On the one hand, as seen in Eq. (13), small singular values $w_n$ correspond to columns of matrices $U$ and $V$ that have very small influence on the matrix of regression coefficients $I_{im}$. On the other hand, these same small singular values have very large impact on the coupling constants in Eq. (14). Therefore, expression (14) should be used for non-maximum values of $n_{\max}$, i.e., by cutting off the small singular values. By that, one removes from the final result those features of the regression coefficients that are poorly defined, and at the same time one keeps only those linear combinations of coupling constants that are well determined by data.

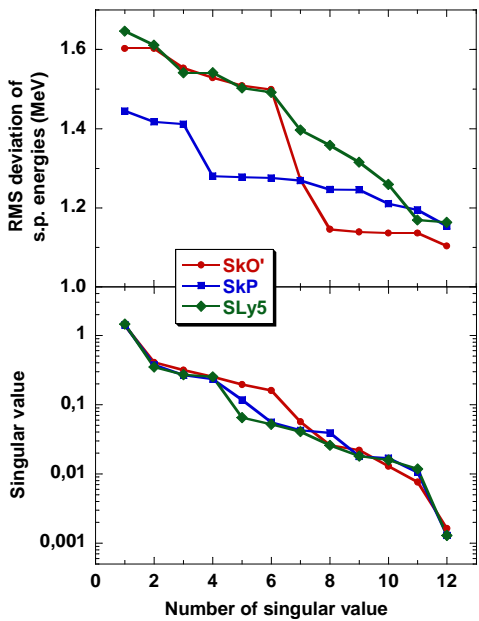

FIG. 15: (Color online) Bottom: singular values corresponding to the regression coefficients $I_{im}$ calculated for the SkO', SkP, and SLy5 Skyrme functionals. Top: the rms deviations of the s.p. energies from experimental data, obtained for a given number of singular values kept in Eq. (14).

In our analysis, for experimental data we adopted $M = 58$ proton and neutron negative s.p. listed in Ref. [19]. Since here we study bare theoretical s.p. energies, we have also subtracted from the experimental values the calculated mass polarization corrections related to the one-body center-of-mass correction [5]. In the upper panel of Fig. 15, we show the rms deviations,

$$\Delta\epsilon_{\rm rms} = \left(\frac{1}{M}\sum_{i=1}^{M}(\Delta\epsilon_i)^2\right)^{1/2}, \qquad (15)$$

of the fitted s.p. energies, obtained for $n_{\max} = 1, 2, \ldots, 12$. For the standard parametrizations studied here, one obtains $\Delta\epsilon_{\rm rms} = 1.73$, 1.61, and 1.45 MeV for SLy5, SkO', and SkP functionals, respectively. At $n_{\max} = 1$, one obtains the rms deviations that are only marginally better. With increasing $n_{\max}$, the results gradually improve, and at $n_{\max} = 12$ (for the full least-square fit) they reach limiting values of about 1.1 MeV.

Although in particular cases, some combinations of the coupling constants are especially important, i.e., at $n_{\max} = 7$–8 for SkO' or at $n_{\max} = 4$ for SkP, the rms deviations decrease rather steadily, and do not show pronounced affects at low values of $n_{\max}$, observed previously for the total binding energies, see Fig. 2 in Ref. [9]. On the one hand, this means that the experimental val-ues of the s.p. energies may, in principle, determine values of more coupling constants than those of the total binding energies. On the other hand, the limiting value of $\Delta\epsilon_{\rm rms} = 1.1$ MeV, which can be obtained by refitting the standard functionals, is rather disappointing. Fig. 16 shows histograms of residuals, Eq. (12), for standard (upper panels) and refitted (lower panels) Skyrme functionals. One can see again that the improvement obtained by the fit is not very spectacular, with quite a number of s.p. energies differing from experiment by 1 MeV or more. This indicates, that to obtain a better agreement with experimental one may have to extend the form of the standard Skyrme functional.

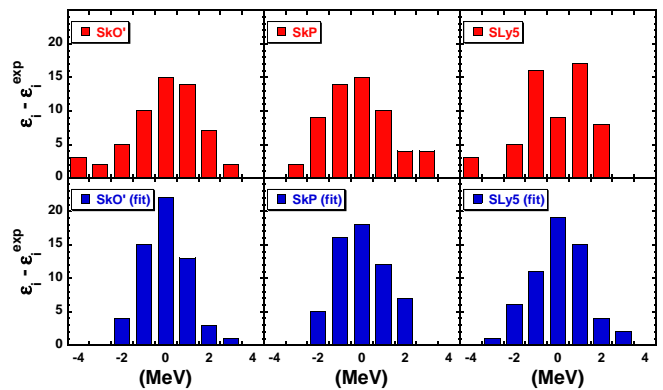

FIG. 16: (Color online) Histograms of residuals, Eq. (12), for standard (upper panels) and refitted (lower panels) Skyrme functionals SkO', SkP, and SLy5.

## IV. CONCLUSIONS

The present work constitutes a step towards the search for a spectroscopic-quality energy density functional, which would properly and precisely account for the single-particle structure of atomic nuclei. We analyzed dependence of single-particle energies on coupling constants of the Skyrme functional and showed that this dependence is almost linear. Such an observation allowed us to focus on linear regression coefficients, which characterize variations of individual single-particle states with changing coupling constants.

We discussed the regression coefficients obtained for single-particle states in 9 doubly magic spherical nuclei. The results turn out to be rather generic, with only a weak dependence on the particular variant of the Skyrme-functional parametrization. We showed that for nuclei near stability, isovector coupling constants have significantly smaller impact on single-particle energies than isoscalar ones. We also showed that the effective-mass coupling constant $C_0^\tau$ cannot be considered as merely changing the overall density of states, which was up to now a commonly accepted view. This coupling constants also significantly influences relative positions of single-particle levels, including the splitting of spin-orbit part-ners. We explained these effects by a detailed analysis of densities and mean fields related to this coupling constant.

By applying the singular value decomposition to the matrix of the regression coefficients, we performed fits of calculated single-particle energies to experimental data. We showed that even by fitting all the 12 coupling constants of the standard Skyrme functional to experiment, one is unable to obtain the rms deviations of single-particle energies below about 1.1 MeV. This discouraging result points out to a necessity of extending the form of the Skyrme functional beyond the standard parametrization.

This work was supported in part by the Academy of Finland and University of Jyväskylä within the FIDIPRO programme and by the Polish Ministry of Science.

[1] J. Dobaczewski, in *Trends in Field Theory Research*, ed. O. Kovras (Nova Science Publishers, New York, 2005) p. 157; nucl-th/0301069.
[2] R.M. Dreizler and E.K.U. Gross, *Density Functional Theory* (Springer, Berlin, 1990).
[3] P. Ring and P. Schuck, *The Nuclear Many-Body Problem* (Springer-Verlag, Berlin, 1980).
[4] W. Kohn and L.J. Sham, Phys. Rev. **140**, A1133 (1965).
[5] M. Zalewski, J. Dobaczewski, W. Satuła, and T.R. Werner, Phys. Rev. C **77**, 024316 (2008).
[6] J. Dobaczewski and J. Dudek, Phys. Rev. **C52**, 1827 (1995); **C55**, 3177(E) (1997).
[7] E. Perlińska, S.G. Rohoziński, J. Dobaczewski, and W. Nazarewicz, Phys. Rev. C **69**, 014316 (2004).
[8] Y.M. Engel, D.M. Brink, K. Goeke, S.J. Krieger, and D. Vautherin, Nucl. Phys. A **249**, 215 (1975).
[9] G.F. Bertsch, B. Sabbey, and M. Uusnäkki, Phys. Rev. C **71**, 054311 (2005).
[10] H. Hellmann, *Einfuehrung in die Quantenchemie* (Deuticke, Leipzig, 1937); R.P. Feynman, Phys. Rev. **56**, 340 (1939).
[11] Ph. Chomaz, K.H.O. Hasnaoui, F. Gulminelli, arXiv:nucl-th/0610027.
[12] S. Weisberg, *Applied linear regression* (Wiley-Interscience, New York, 2005).
[13] K. Bennaceur and J. Dobaczewski, Comput. Phys. Commun. **168**, 96 (2005).
[14] E. Chabanat, P. Bonche, P. Haensel, J. Meyer, and R. Schaeffer, Nucl. Phys. **A627** (1997) 710; **A635** (1998) 231.
[15] J. Dobaczewski, H. Flocard and J. Treiner, Nucl. Phys. **A422**, 103 (1984).
[16] P.-G. Reinhard, D.J. Dean, W. Nazarewicz, J. Dobaczewski, J.A. Maruhn, and M.R. Strayer, Phys. Rev. **C60**, 014316 (1999).
[17] W. Satuła, R.A. Wyss, and M. Zalewski, arXiv:0802.3605.
[18] W.H. Press, B.P. Flannery, S.A. Teukolsky, and W.T. Vetterling, "Singular Value Decomposition", §2.6 in *Numerical Recipes in FORTRAN: The Art of Scientific Computing* (Cambridge University Press, Cambridge, 1992).
[19] N. Schwiertz, I. Wiedenhover, and A. Volya, arXiv:0709.3525.